\documentclass[conference]{IEEEtran}

\IEEEoverridecommandlockouts
\usepackage{cite,xurl}
\usepackage{amsmath,amssymb,amsfonts}
\usepackage{algorithmic}
\usepackage{graphicx}
\usepackage{textcomp}
\usepackage{algorithm2e}
\usepackage{xcolor}
\def\BibTeX{{\rm B\kern-.05em{\sc i\kern-.025em b}\kern-.08em
    T\kern-.1667em\lower.7ex\hbox{E}\kern-.125emX}}

\usepackage{caption}
\usepackage{subcaption}
    
\usepackage{tikz}
\usetikzlibrary{quantikz}

\usepackage{subcaption}

\begin{document}

\title{On Distributed Quantum Computing with Distributed Fan-Out Operations}

\author{\IEEEauthorblockN{Seng W. Loke}
\IEEEauthorblockA{\textit{School of Information Technology, Deakin University, Burwood, VIC 3125, Australia.} \\
seng.loke@deakin.edu.au}
 }


\maketitle

\begin{abstract}
We compare different circuits implementing distributed versions of quantum computations, using entangled pairs only, and using distributed fan-out operations (using GHZ states). We highlight the advantages of using distributed fan-out operations in terms of reductions in circuit depth and (possibly) entanglement resources. We note that distributed fan-out operations (or notably, distributed GHZ states) could be a ``primitive'' building block for distributed quantum operations in the same way as entangled pairs are, if distributed GHZ states could be realized efficiently. 
\end{abstract}


\section{Introduction}
There has been increasing interest in distributed quantum computing in industry\footnote{For example, see ~\cite{photonics}, \url{https://spectrum.ieee.org/quantum-computers}, \url{https://newsroom.ibm.com/2025-11-20-ibm-and-cisco-announce-plans-to-build-a-network-of-large-scale,-fault-tolerant-quantum-computers}} and academia~\cite{BARRAL2025100747,knörzer2025,CALEFFI2024110672,arquin}.\footnote{See also \url{https://quantnet.lbl.gov} and \url{https://www.ox.ac.uk/news/2025-02-06-first-distributed-quantum-algorithm-brings-quantum-supercomputers-closer}} Much work has focused on the use of Bell pairs for realizing distributed CNOT gates, as such gates together with single qubit gates provide a universal gate set for quantum computations, and creation of entangled pairs would be a well-known form of multi(two)-qubit entanglement. 

Previous work advocated the use of {\em fan-out} operations, with single control qubits and multiple target qubits, amenable to efficient implementation in some types of quantum hardware such as trapped ion quantum computers, which in turn, enables efficient (or reduced depth) quantum computations, such as for SWAP test, shared control Toffoli operations, Hadamard test and to improve the implementation of quantum memory architectures~\cite{fenner23,hoyer05}. However, they do not consider distributed quantum computations as we do here. 

Distributed fan-out using distributed GHZ states have been proposed in~\cite{Yimsiriwattana:2004xhy}.
This paper discusses further uses of the distributed fan-out operation for distributed quantum computing, including its generality and efficiency advantages for distributed quantum computations.

In the rest of this paper, we first describe distributed fan-out operations and then look at examples where its use can be advantages, including distributed quantum fourier transforms and implementation of general $n$-ary unitaries. We  conclude with avenues for future work.

\section{Distributed Fan-Out Operations}

\begin{figure*}
\centering
  \begin{tabular}{c c}
      \begin{minipage}[c]{0.8\textwidth}
        \centering
  \includegraphics[width=0.8\textwidth]{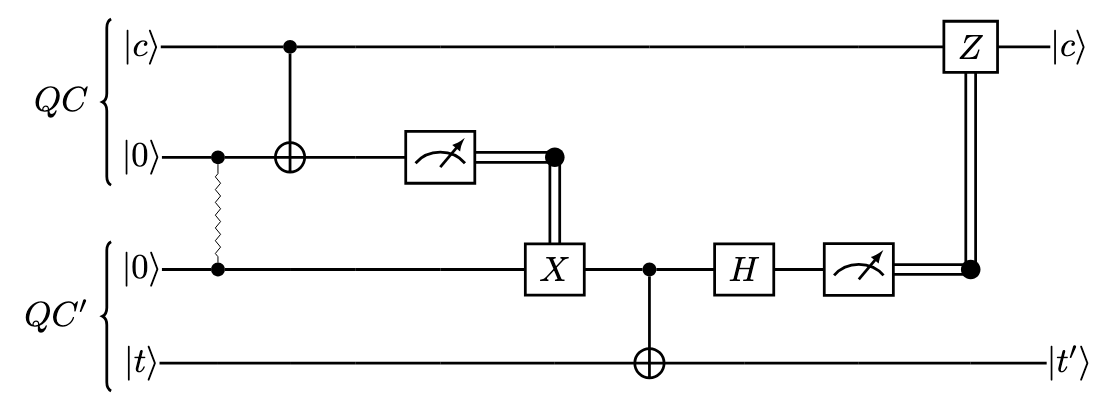} 
    \end{minipage}     
    & $\equiv$ 
\end{tabular}
\begin{quantikz}  c~ & \ctrl{1}\gategroup[2,steps=1,style={dashed,inner sep=6pt}]{Dc-U}   & \qw  \\ 
t  & \gate{U}   & \qw    \end{quantikz} 
  \caption{Distributed control-$U$ (i.e., dCNOT is when U=$X$ ($\oplus$) as shown)   between nodes $QC$ and $QC'$ (the {\em computation} qubits are $c$, the control qubit, and $t$, the target qubit), and the wavy line illustrates a Bell pair  involving  the  {\em communication} qubits from the two nodes both initially $\ket{0}$. The right hand side shows the notation we will use for a distributed controlled-$U$ operation.}
\label{dcnot}
\end{figure*}

We first consider the distributed CNOT gate (or  dCNOT, in short), which is needed as a resource for each dCNOT gate, as shown in Figure~\ref{dcnot}.

\begin{figure*}

\begin{center}
\scalebox{.9}{
      \begin{quantikz} 
   \lstick[2]{A'}    c~                           \qw   & \qw  & \ctrl{1}  & \qw   & \qw & \qw & \qw &   \qw &  \qw   &   \gate{Z^{m_1 \oplus m_2 \oplus m_3}} & \qw c  \\
    a_0    \qw & \ctrl{}{}  & \targ{}  & \qw & \meter{} &  \cwbend{5} &    m_0      &            &  &         \\
    \lstick[2]{A}    a_1    \qw & \ctrl{}{}  & \qw       & \qw & \qw      & \gate{X^{m_0}}   & \ctrl{1} & \gate{H}& \meter{} &  \cwbend{-2} &  m_1 \\
      t1   \qw & \qw        & \qw       & \qw & \qw      & \qw         & \gate{U_1}       & \qw        & \qw &  \qw &\qw   \\
          \lstick[2]{B}    a_2   \qw & \ctrl{}{}  & \qw       & \qw & \qw      & \gate{X^{m_0}}   & \ctrl{1} & \gate{H}  & \meter{} &  \cwbend{-2}   & m_2    \\
      t2   \qw & \qw        & \qw       & \qw & \qw      & \qw         & \gate{U_2}       & \qw        & \qw &  \qw &\qw  \\
                \lstick[2]{C}    a_3   \qw & \ctrl{}{}  & \qw       & \qw & \qw      & \gate{X^{m_0}}   & \ctrl{1} & \gate{H}  & \meter{} &  \cwbend{-2}   & m_3    \\
      t3   \qw & \qw        & \qw       & \qw & \qw      & \qw         & \gate{U_3}       & \qw        & \qw &  \qw &\qw  
         \arrow[from=2-2,to=4-2,squiggly,dash,line width=0.1mm]{}
         \arrow[from=7-2,to=4-2,squiggly,dash,line width=0.1mm]{}
    \end{quantikz}
    }
    $\equiv$
    \begin{quantikz}
 c~ & \ctrl{3}\gategroup[4,steps=1,style={dashed,inner sep=6pt}]{Dfan-out}   & \qw  \\
 t_1  & \gate{U_1}   & \qw \\ 
  t_2  & \gate{U_2}   & \qw \\ 
  t_3  & \gate{U_3}   & \qw 
\end{quantikz}
\end{center}
\caption{Distributed fan-out operation with single control qubit (on $A'$) for multiple target qubits (one on $A$, one on $B$ and one on $C$) - all target qubits on different nodes from the control qubit. The right hand side shows the notation we will use for distributed fan-out throughout the paper.}
\label{dfanout}
\end{figure*}
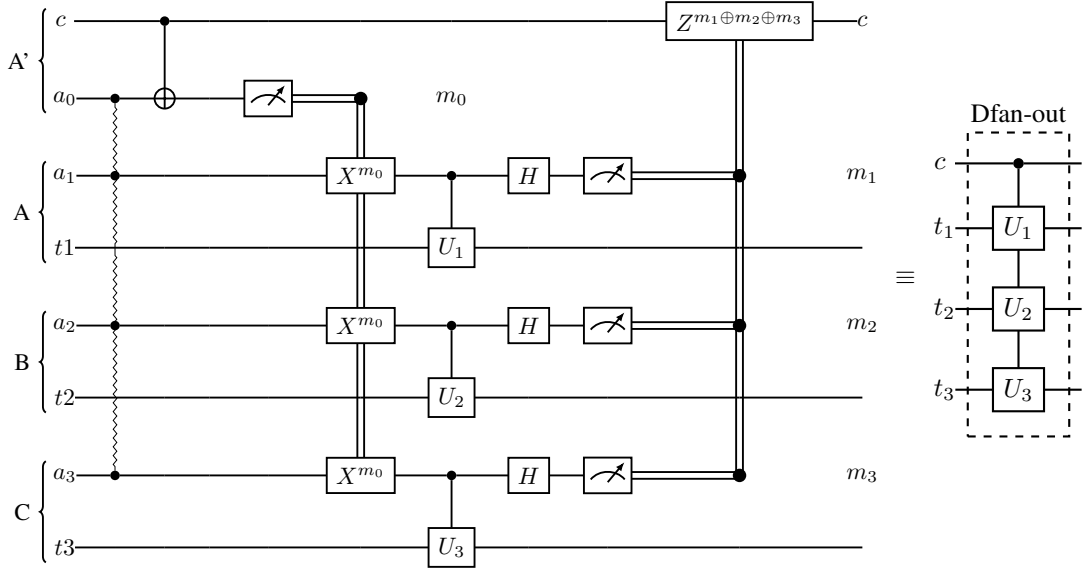

By a distributed fan-out  operation, we refer to an operation where the same control qubit is used for multiple target qubits, on different nodes. This can arise from the structure of a quantum circuit itself, or from a control-$U$ operation where the unitary $U$ spans multiple qubits and is decomposed into operations executed on qubits distributed on different nodes~\cite{Yimsiriwattana:2004xhy,loke2023distributed,PhysRevA.107.L060601}. For example, the quantum circuit such as:
    \begin{center}
          \begin{quantikz}
     c~     & \ctrl{1} & \ctrl{2}  & \ctrl{3}  & \qw       \\
     t_1     & \gate{U_1} &  \qw  & \qw  & \qw      \\
     t_2     & \qw        &  \gate{U_2}   & \qw   & \qw    \\
     t_3      & \qw     &  \qw  & \gate{U_3}  & \qw             \\
    \end{quantikz}
    \end{center}
i.e., a {\em multitarget control-$U$},   where $U_1$ acts on qubit~$t1$, $U_2$ acts on qubit~$t2$, and $U_3$ acts on qubit~$t3$, each qubit on a different node, will result in the distributed operation as shown in Figure~\ref{dfanout}; note that the wavy line in the figure connecting the four (communication) qubits $a_0,a_1,a_2,a_3$  represents a distributed 4-qubit GHZ state of the form $\frac{1}{\sqrt{2}}(\ket{0 0 0 0}+\ket{1111})_{a_0 a_1 a_2 a_3}$, over four nodes $A'$, $A$, $B$ and $C$.

\section{Examples}
We look at some uses of the distributed fan-out operation for reduced depth distributed quantum computations.

\subsection{Distributed Quantum Fourier Transform (QFT)}

Consider the Quantum Fourier Transform (QFT) algorithm over $n$ qubits and let $N= 2^n$, that computes:\footnote{See \url{https://pennylane.ai/qml/demos/tutorial_qft}}
\[
QFT \ket{x} = \bigotimes_{k=n-1}^{0} \Big(\ket{0} + exp(\frac{2\pi i 2^k}{2^n} x)\ket{1}\Big) 
\]
where $x \in \{0,\ldots,N-1 \}$.

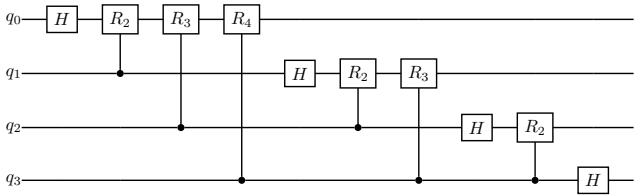
\begin{figure}

\begin{center}
\scalebox{.67}{
      \begin{quantikz}
     q_0 & \gate{H}    & \gate{R_2} & \gate{R_3}  & \gate{R_4}  & \qw & \qw &  \qw & \qw & \qw & \qw &\qw\\
     q_1 & \qw  & \ctrl{-1} &  \qw  & \qw  & \gate{H}    & \gate{R_2} & \gate{R_3} & \qw & \qw & \qw & \qw\\
     q_2 & \qw    & \qw &         \ctrl{-2}   & \qw   & \qw & \ctrl{-1} &  \qw & \gate{H} & \gate{R_2} & \qw &\qw\\
     q_3 &  \qw   & \qw     &  \qw  & \ctrl{-3}  & \qw    &  \qw & \ctrl{-2}  & \qw & \ctrl{-1} & \gate{H}  & \qw   \\
    \end{quantikz}
    }
    \end{center}
\caption{Quantum Fourier Transform with 4 qubits using controlled rotations.}
\label{qft4-1}
\end{figure}

One formulation is shown in Figure~\ref{qft4-1}, where
\[
R_k=
\begin{bmatrix}
    1 & 0 \\
    0 & exp(\frac{2\pi i}{2^k}) 
\end{bmatrix}
\]

\begin{figure}

\begin{center}
\scalebox{.57}{
      \begin{quantikz}
     q_0 & \gate{H}    & \ctrl{1} & \ctrl{2}  & \ctrl{3}  & \qw & \qw &  \qw & \qw & \qw & \qw &\qw\\
     q_1 & \qw  & \gate{P(\frac{\pi}{2})} &  \qw  & \qw  & \gate{H}    & \ctrl{1} & \ctrl{2} & \qw & \qw & \qw & \qw\\
     q_2 & \qw    & \qw &   \gate{P(\frac{\pi}{4})}   & \qw   & \qw & \gate{P(\frac{\pi}{2})} &  \qw & \gate{H} & \ctrl{1} & \qw &\qw\\
     q_3 &  \qw   & \qw     &  \qw  & \gate{P(\frac{\pi}{8})}  & \qw    &  \qw & \gate{P(\frac{\pi}{4})}  & \qw & \gate{P(\frac{\pi}{2})} & \gate{H}  & \qw   \\
    \end{quantikz}
    }
    \end{center}
\caption{Quantum Fourier Transform with 4 qubits using controlled phase operations.}
\label{qft4-2}
\end{figure}
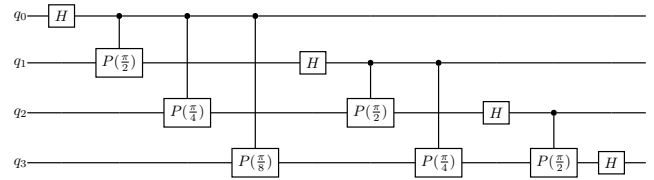

But we use the formulation from~\cite{PRXQuantum.4.040318}, using controlled-phase gates shown in Figure~\ref{qft4-2}, where 
\[
P(\theta)=
\begin{bmatrix}
    1 & 0 \\
    0 & e^{i\theta} 
\end{bmatrix}
\]
which shows the fan-out operations.

\begin{figure}
\centering
\scalebox{0.75}{
    \begin{quantikz}
 q_0~  & \gate{H} & \ctrl{3}\gategroup[4,steps=1,style={dashed,inner sep=6pt}]{Dfan-out}   & \qw & \qw & \qw& \qw & \qw & \qw & \qw & \qw\\
 q_1  & \qw & \gate{P(\frac{\pi}{2})}   & \qw & \gate{H} & \ctrl{2}\gategroup[3,steps=1,style={dashed,inner sep=6pt}]{Dfan-out} & \qw & \qw & \qw & \qw & \qw\\ 
  q_2 & \qw & \gate{P(\frac{\pi}{4})}    & \qw & \qw & \gate{P(\frac{\pi}{2})} & \qw & \gate{H} & \ctrl{1}\gategroup[2,steps=1,style={dashed,inner sep=6pt}]{Dc-U}   & \qw  & \qw \\ 
  q_3 & \qw & \gate{P(\frac{\pi}{8})}   & \qw & \qw & \gate{P(\frac{\pi}{4})} & \qw &   \qw & \gate{P(\frac{\pi}{2})} & \gate{H} & \qw 
\end{quantikz}
}
\caption{Distributed QFT over 4 qubits on 4 nodes (one qubit per node), ignoring bit reversal operations.}
\label{dqft-cp}
\end{figure}
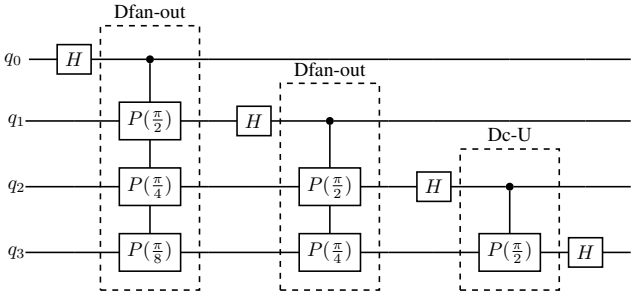

Given the fan-out operations, we can then implement a dsitributed version of the quantum Fourier transform (for 4 qubits) using the distributed fan-out and distributed controlled-$U$ operations as shown in Figure~\ref{dqft-cp}.

Distributed QFT (dQFT) over four nodes, one qubit per node, can be realized using fan-out operations, i.e., one for the first block of three controlled-$P(\theta)$ operations using a 4-qubit GHZ state as in Figure~\ref{dfanout}, and another for the next block of two controlled-$P(\theta)$ operations using a 3-qubit GHZ state (qubits $q_1$,$q_2$ and $q_3$). The last controlled-$P(\theta)$ operations can be done via a Bell pair and a circuit similar to that in Figure~\ref{dcnot}.

One can also use 6 Bell pairs, converting each controlled-$P(\theta)$ operation into a distributed control operation similar to Figure~\ref{dcnot}. 

In general, using only Bell pairs, for an $n$-qubit dQFT over $n$ nodes, one qubit per node, we will need $n(n-1)/2$ Bell pairs.
Using fan-out operations, we would need $(n-2)$ GHZ states of different sizes (i.e., one $n$-qubit GHZ state, one $(n-1)$-qubit GHZ state, one $(n-2)$-qubit GHZ state,... and one $3$-qubit GHZ state), and one Bell pair. If a distributed GHZ state can be done in ``one step'' (or one shot), then the depth is significantly reduced~\cite{Ainley:2024bdu,singh26,PhysRevA.107.L060601}. Otherwise, each $k$-qubit GHZ state can be formed using $log~k$ steps using $k$ Bell pairs (with parallelism using a tree-like approach), retaining a reduced circuit depth (e.g., $O(n~log~n)$ steps), even if not a reduction in Bell pairs required.

\subsection{Distributed Quantum Approximate Optimization Algorithm (QAOA)}

Fan-out operations might be useful for quantum circuits used in computations in applications of the QAOA~\cite{farhi14}.
For example, applying to a max-cut problem for a graph $G$, a cost Hamiltonian might be of the form:
\[
\mathcal{H}_c = \sum_{(p,q)\in G} k_{pq} Z_p Z_q
\]
for some constants $k_{pq}$, and an  operation to compute repeatedly includes one of the form $e^{- i \gamma \mathcal{H}_c}$ involving the cost Hamiltonian, together with an operation for the mixing Hamiltonian, for some parameter $\gamma$  possibly different each time, using the Suzuki-Trotter scheme\footnote{For example, see \url{https://pennylane.ai/qml/demos/tutorial_qaoa_maxcut}}.

Effectively, we need a circuit to compute a unitary of the form:
\[
e^{-i  \sum_{(p,q) \in G} \theta_{pq} Z_p Z_q} = \Pi_{(p,q) \in G}~ e^{-i  \theta_{pq} Z_p Z_q}
\]

Noting that
\[
e^{-i \theta Z_p Z_q} = e^{i \theta}\, \Bigl(R_z^{(p)}(2\theta) \otimes R_z^{(q)}(2\theta) \Bigr)\,\mathrm{CP}_{pq}\!\left(-4\theta\right)
\]
where $CP_{pq}$ is a controlled-phase operation:
\[
CP_{pq}(\alpha) = \ket{0}\!\bra{0} \otimes I +  \ket{1}\!\bra{1} \otimes P(\alpha)
\]
where $P(\alpha) = \ket{0}\!\bra{0} + e^{i\alpha}\ket{1}\!\bra{1}$, and
\[
R_Z(\theta)=
\begin{bmatrix}
    e^{-i\theta} & 0 \\
    0 & e^{i\theta} 
\end{bmatrix}
\]

\begin{figure}
\centering
\begin{quantikz}
 p~ & \ctrl{3}\gategroup[4,steps=1,style={dashed,inner sep=6pt}]{Dfan-out}  & \gate{R_Z(2\theta_0+2\theta_1+2\theta_2)} & \qw  \\
 q  & \gate{P(-4\theta_0)}  & \gate{R_Z(2\theta_0)}  & \qw \\ 
q' & \gate{P(-4\theta_1)}   & \gate{R_Z(2\theta_1)} & \qw \\ 
q''  & \gate{P(-4\theta_2)}  & \gate{R_Z(2\theta_2)}   & \qw 
\end{quantikz}
\caption{Distributed form for the example circuit for $G_{ex}$}
\label{dqaoa}
\end{figure}
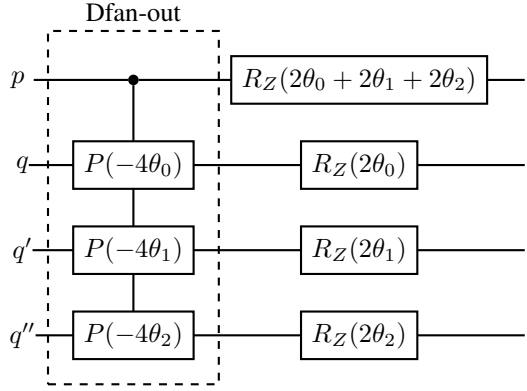

Then, the circuit for the operator $e^{-i  \theta_{pq} Z_p Z_q}$ takes the form, ignoring global phases:
    \begin{center}
          \begin{quantikz}
     p     & \ctrl{1} & \gate{R_Z(2\theta)}    & \qw       \\
     q     & \gate{P(-4\theta)} &  \gate{R_Z(2\theta)}      & \qw      \\ 
    \end{quantikz}
    \end{center}
Now, suppose an example graph $G_{ex}$ is such that we have a product:
\[
e^{-i \theta_{2} Z_p Z_{q''}} e^{-i \theta_{1} Z_p Z_{q'}} e^{-i \theta_{0} Z_p Z_{q}}
\]
where $\theta_0 = \theta_{pq}$, $\theta_1 = \theta_{pq'}$, and $\theta_2 = \theta_{pq''}$, that is, there are  edges from $p$ to $q$, $q'$ and $q''$ (or $(p,q),(p,q'),(p,q'') \in G_{ex}$), then, we have the circuit:

    \begin{center}
    \scalebox{0.6}{
          \begin{quantikz}
     p     & \ctrl{1} & \gate{R_Z(2\theta_0)}    & \ctrl{2} & \gate{R_Z(2\theta_1)}  & \ctrl{3} & \gate{R_Z(2\theta_2)} & \qw      \\
     q     & \gate{P(-4\theta_0)} &  \gate{R_Z(2\theta_0)}   &  \qw & \qw     & \qw   &\qw & \qw     \\ 
    q'     & \qw & \qw  & \gate{P(-4\theta_1)} & \gate{R_Z(2\theta_1)}    & \qw     & \qw & \qw   \\
     q''    & \qw & \qw    & \qw &  \qw &\gate{P(-4\theta_2)} & \gate{R_Z(2\theta_2)}    & \qw      \\ 
    \end{quantikz}
    }
    \end{center}
    where we can move the $R_Z$ operations to the end (and also combine them) to get:
        \begin{center}
           \scalebox{0.75}{
          \begin{quantikz}
     p     & \ctrl{1}   & \ctrl{2} & \qw  & \ctrl{3}  & \gate{R_Z(2\theta_0+2\theta_1+2\theta_2)}       & \qw \\
     q       & \gate{P(-4\theta_0)}  & \qw     & \qw   &\qw  & \gate{R_Z(2\theta_0)} & \qw   \\ 
    q'     & \qw  & \gate{P(-4\theta_1)}     & \qw     & \qw   & \gate{R_Z(2\theta_1)} & \qw\\
     q''    & \qw & \qw   &  \qw &\gate{P(-4\theta_2)}   & \gate{R_Z(2\theta_2)}     &\qw   \\ 
    \end{quantikz}
    }
    \end{center}
    We can then implement this using distributed fan-out as illustrated in Figure~\ref{dqaoa}.

That is, if the qubits were distributed over four nodes, one qubit per node, we can then apply the distributed fan-out operation similar to  Figure~\ref{dfanout} to reduce circuit depth. 
This will not be advantages for all graphs but for graphs with structure, where there are many one to many connectivity. For such graphs, advantages can be significant given the number of  operations required in a Suzuki-Trotter scheme, but also due to  the potentially large number of (re-)executions of the circuits to gather statistics.


\section{General Quantum Circuits}
Given that any $n$-qubit unitary operation can be written as
\[
U \;=\; e^{\left(
  -i \sum_{\alpha} c_{\alpha} \, P_{\alpha} \right)},
\quad
P_{\alpha} \in \{ I, X, Y, Z \}^{\otimes n},
\quad
c_{\alpha} \in \mathbb{R}.
\]

\begin{figure}
   \begin{center}
   \scalebox{0.95}{
          \begin{quantikz}
     a     & \qw \oplus & \qw \oplus  & \qw \oplus  & \qw       \\
     q_1     & \ctrl{-1} &  \qw  & \qw  & \qw      \\
     q_2     & \qw        &  \ctrl{-2}   & \qw   & \qw    \\
     q_3      & \qw     &  \qw  & \ctrl{-3}  & \qw             \\
    \end{quantikz}
    }
    $\equiv$
    \scalebox{0.8}{
          \begin{quantikz}
     a     & \gate{H} & \ctrl{1} & \ctrl{2}  & \ctrl{3}  & \qw    & \gate{H}   & \qw  \\
     q_1     & \gate{H} & \qw \oplus &  \qw  & \qw  & \qw    & \gate{H} & \qw   \\
     q_2    & \gate{H}  &\qw     &  \qw \oplus      & \qw  & \qw  & \gate{H} & \qw \\
     q_3      &    \gate{H} &  \qw  & \qw & \qw \oplus   & \qw    & \gate{H} & \qw         \\
    \end{quantikz}
    }
    \end{center}
    \caption{Parity operation over three qubits $q_1$, $q_2$, and $q_3$ computed on ancilla $a$, expressed using fan-out. }
    \label{parity}
    \end{figure}
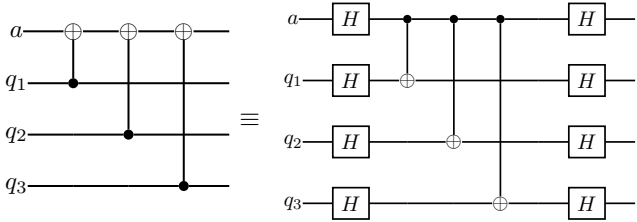

To implement, say, one component $e^{-i c_\alpha P_{\alpha}}$, we can use the following approach.
First consider implementing a unitary of the form:
\[
e^{-i\theta Z_1...Z_n} 
\]
When applied to a computational basis state $\ket{x}$=$\ket{x_1...x_n}$, we get
\[
e^{-i\theta Z_1...Z_n} \ket{x} = e^{-i\theta (-1)^{parity(x)}}  \ket{x}
\]
where $parity(x) = x_1 \oplus ... \oplus x_n$.
We can compute the parity using a distributed parity gate, implemented via  a fan-out operation due to an equivalence as illustrated in Figure~\ref{parity} with $n=3$, with qubits $q_1$, $q_2$ and $q_3$~\cite{moore99}.

Using the parity gate, we can compute the parity in an  ancilla qubit $a$, and then apply the rotation $R_Z(2\theta)$ to the ancilla, and then apply the parity gate again to the ancilla (to uncompute), and the outcome is $e^{-i\theta Z_1 Z_2 Z_3} \ket{\psi}_{q_1 q_2 q_3}$, where $\ket{\psi}_{q_1 q_2 q_3}$ is the state of the three qubits.
Starting with an ancilla qubit and the three qubits:
\[
\ket{0}_a \otimes \sum_{q_i \in \{0,1\}} \alpha_{q_1 q_2 q_3} \ket{q_1q_2 q_3}
\]
we apply the parity gate to obtain:
\[
 \sum_{q_i \in \{0,1\}} \ket{q_1 \oplus q_2 \oplus q_3}_a \otimes \alpha_{q_1 q_2 q_3}  \ket{q_1q_2 q_3}
\]
Then, we apply the rotation $R_Z(2\theta)$ to the ancilla to get:
\[
\sum_{q_i \in \{0,1\}}  e^{-i\theta(-1)^{q_1\oplus q_2\oplus q_3}} \ket{q_1 \oplus q_2 \oplus q_3}_a \otimes  \alpha_{q_1 q_2 q_3} \ket{q_1q_2,q_3}
\]
And then applying the parity gate on the qubits and ancilla to uncompute, we get:
\begin{align*}
& \ket{0}_a \otimes \sum_{q_i \in \{0,1\}}  e^{-i\theta(-1)^{q_1\oplus q_2\oplus q_3}}   \alpha_{q_1 q_2 q_3} \ket{q_1q_2,q_3} \\
&= \ket{0}_a \otimes e^{-i\theta Z_1...Z_n} \ket{\psi}_{q_1 q_2 q_3}
\end{align*}
which is what we wanted.
We can use this to implement $e^{-i c_{\alpha} P_{\alpha}}$ for any Pauli string $P_{\alpha}$ by using equivalences such as $HR_Z(\theta)H=R_X(\theta)$. For example,
$e^{-i\theta XZXZ}$ can be realized by  computing $e^{-i\theta ZZZZ}$ preceded and succeeded by $H$ gates for the first and third qubits.

With one qubit per node, $n$ qubits distributed over $n$ nodes, we can use a distributed $n$-qubit parity gate implemented using the distributed $n$-qubit fanout operation and $H$ gates due to the equivalence above. We can  
generalize to $n$ qubits with constant depth with the use of distributed GHZ states since the scheme for GHZ states (Figure~\ref{dfanout}) has constant depth for $n$ qubits, provided the distributed $n$-ary parity gate has constant depth (not increasing with $n$)!

\begin{figure*}[ht]
\centering
\begin{quantikz}
& & \gate[5]{\mbox{single qubit gates}} & \gate[5]{Dfan-out} & \gate[5]{\mbox{single qubit gates}} & \gate[5]{Dfan-out} & \dots & \gate[5]{\mbox{...}} &\qw \\
&  &&& &&  \ldots & & \qw \\
& &&& && \ldots & &\qw \\
\wave &&&&&& \ldots & &\qw \\
&&&&&& \ldots & & \qw 
\end{quantikz}
\caption{An illustration of a layered view of a typical distributed quantum circuit interleaving single qubit (local) gates and distributed fan-out operations (of course, not all distributed fan-out operations will always involve all qubits).}
    \label{layers}
\end{figure*}
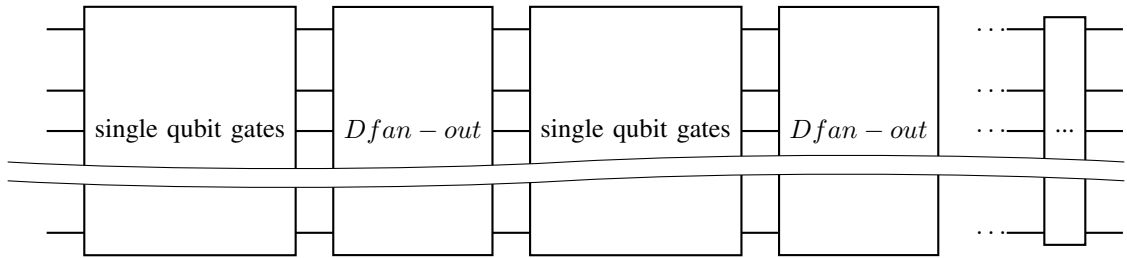

In general, the distributed computation of a general $n$-qubit unitary gate can be done in layers of single qubit gates (all local) and distributed fan-out operations, as illustrated in Figure~\ref{layers}.

Single qubit gates with fan-out gates are universal for quantum circuits, clearly since a CNOT gate is just a fan-out gate with one target.

Hence, for such an implementation of distributed $n$-qubit unitary operations over $n$ nodes, one qubit per node, if the distributed fan-out gates (or parity gates) have constant depth $D_{Dfan-out}$, then a circuit with $k$ layers of single qubit gates interleaved with $m$ layers of distributed fan-out gates will have depth $O(k+m~D_{Dfan-out})$. If a distributed $r$-qubit GHZ state is realized using Bell pairs connected in a tree-like structure~\cite{Yimsiriwattana:2004xhy}, then $log~r$ steps with $r$ Bell pairs might be used, that is, the circuit depth might have $O(k+m~log(n))$ circuit depth. The argument for scalability here is that as $n$ increases, we maintain a relatively lower (possibly constant) circuit depth using distributed fan-outs, as opposed to the situation when using just sequences of Bell pairs (e.g., $O(k+mn)$ in such cases).

\section{Conclusion}
We have outlined in this paper the advantages and generality of the distributed fan-out operation for distributed quantum computing. Despite higher costs, with large enough $n$, arguably, one can have much reduced circuit depth compared to using Bell pairs alone. For large enough $n$, this can help maintain the high relative performance advantage of quantum operations compared to exponential time classical simulations, even when distributed qubits are used, despite the higher costs  of entangling qubits across nodes. 
Also, future distributed quantum computing systems  can be optimized to deliver high speeds for forming $k$-qubit GHZ states as $k$ increases (and caching  such GHZ states), to support efficient ``unbounded'' distributed fan-out operations. While we have assumed one qubit per node in our examples, a similar analysis can be done for multiple qubits per node.
Also, gate reordering approaches to create larger fan-out operations (or gate packets as called in~\cite{Mengoni:2025lsk}) can help further  reduce circuit depth.

\bibliographystyle{plain}
\bibliography{refs-new}
\end{document}